# AUTOMATED CHECKING AND VISUALIZATION OF INTERLOCKS IN THE ISAC CONTROL SYSTEM

R. Keitel and R. Nussbaumer, TRIUMF, Vancouver, B.C., V6T 2A3, Canada

## Abstract

The EPICS based control system of the ISAC radioactive beam facility supervises several sub-systems, which are controlled by PLCs. Most of the devices are protected by non-trivial interlocks, which are implemented with ladder-logic software. Detailed information on interlock state and the individual interlock conditions are accessible for each device at the EPICS operator interface level. With the increasing number of ISAC devices, the interactive generation and maintenance of these displays with standard EPICS tools was too labor-intensive and error-prone. Software was developed, which uses a printout of the PLC program as well as reports from the ISAC device database to a) check the interlock implementation in the PLC against interlock specifications in natural language and b) generate device displays with a graphical representation of interlock state and details of interlock conditions.

## 1 BACKGROUND

The control system of the ISAC Radioactive Beam Facility at TRIUMF is implemented using the EPICS toolkit [1,2]. Control of the vacuum sub-systems and the sensitive target ion source sub-systems is implemented with Modicon Quantum series PLCs, in order to maintain close to 100% up-time for these systems. The PLCs are peer nodes on the controls Ethernet and are supervised by the EPICS input/output computers (IOCs) using EPICS TCP/IP drivers [1].

Most of the devices in these sub-systems have non-trivial interlocks, which are implemented in the ladder logic of the PLC program. The detailed interlock state of each device can be displayed on the operator console by calling up device control panels from named buttons on synoptic sub-system displays.

Initially, these device control panels were generated interactively using the EPICS *edd* display-editing tool. The compliance of the ladder logic implementation with the interlock specifications was checked by formal walk-throughs of the ladder code. With the number of ISAC devices growing quickly, these methods proved to be quite unsatisfactory and error-prone.

A set of Perl tools was developed, to automate both the interlock checking and interlock visualization processes. It should be noted, that the approach presented here is not a general solution to the problem, but relies on some ladder programming conventions adopted at ISAC and also on some features of the MODSOFT ladder programming software.

## 2 ISAC INTERLOCKS

### 2.1 Interlock Specification

The ISAC controls software group realized early in the project, that interlock specifications could be obtained from the equipment specialists much faster, if the controls group provided draft documents for mark-up by the specialist. Interlocks specifications are entered as device instance data in the ISAC relational device database (RDB). The database contains logical expressions for "on" interlocks, "off" interlocks, and "trips". The logical expressions do not use control system variables, but express conditions in natural language, so that they make intuitive sense to the equipment specialists. In order to match the PLC implementation (see below), a maximum number of eight primary conditions are allowed. If this is not enough, placeholder names are used for the primary condition, which are then detailed separately. An example of an interlock specification from the ISAC device database is shown in Fig. 1.

Interlock specification documents are generated for each sub-system as reports from the RDB.

---

Interlock to turn on:
(DTL:CG1A < HEBT:CG2 OR DTL:CG1A < 50 mTorr)
AND (HEBT:CG2 < 150 mTorr OR rough mode)
Trip:
none

rough mode: HEBT:IV0 closed AND HEBT:TP2 off AND DTL:PV1 open

---

Figure 1: Interlocks specification from the ISAC device database

## 2.2 Interlock Implementation

ISAC uses Modicon PLCs of the Quantum series. These are programmed in Ladder Logic using the Modsoft programming package. PLC devices are interfaced to the supervisory EPICS systems with soft status and command registers. The minimum interface is a 16-bit wide status register of which one bit is an "interlock ok summary" status bit (STATINT) and 8 bits are reserved for detailed interlock status, i.e. "interlock 1 bad" (BAD1), "interlock 2 bad" (BAD2), etc. As a convention, the ladder implementation of the interlock ok summary uses only the BAD1.. BAD8 bits. In addition, the device "on" and "off" status bits (STATON, STATOFF) are used to distinguish "on" conditions, "off" conditions, and "trip" conditions.

Figure 2 shows a screen capture from the Modsoft ladder editor displaying the ladder logic implementation of the specification from Fig. 1.

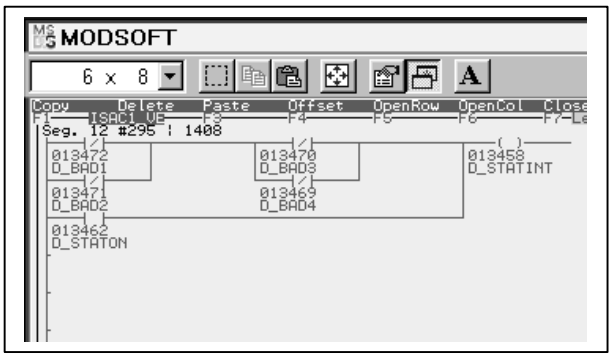

Figure 2: Ladder logic implementation of the interlock summary specification from Fig. 1, containing on "on" interlocks

## 2.3 Interlock Checking

The ISAC PLC programs are implemented using the visual ladder editor of the Modsoft programming package. Although the binary program format is proprietary, the Modsoft programming package can print the program content to a file using (pretty ugly-looking) ASCII character graphics.

In order to automate the checking of the interlock specification against the PLC program implementation
- Code was added to the ISAC device RDB application to report the interlock specification to an ASCII meta-file. For simplicity, any interlocks more complex than a series of ANDed nodes consisting of ORed primitive conditions were flagged as complex and referred to manual checking.
- A Perl tool was developed which analyses the PLC ladder program. The tool generates an ASCII meta-file in the same format produced by the RDB and reports all discrepancies between the ladder program and the interlock specification.

With this - admittedly quite simplistic - approach, the interlocks of 97% of the ISAC vacuum and ion source system devices could be verified to be correctly implemented. The remaining devices are checked by code inspection. The interaction of the different tools is shown in Figure 3.

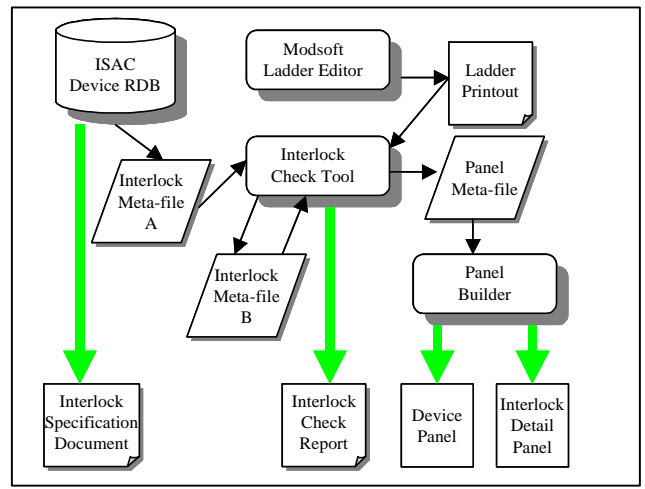

Figure 3: Interaction of RDB and tools

## 2.4 Interlock Visualization

The presentation of the device interlocks to the ISAC operators evolved in three stages.

*a) Interactive generation:* Initially, panels were created with the EPICS edd display editor from parameterized display pages, using cut-and-paste. Then the logic was drawn and the interlock condition texts were added according to the interlock specification. This stage established the general look and feel of the device control panels. It was decided to display the logic with flow lines like in a ladder program, changing the colour of the condition texts depending on the ok/bad state of the interlocks. In order to make better use of screen real estate, the flow lines were flipped to run top-to-bottom on the panels instead of the left-to-right direction in the ladder program.

b) *Automatic generation from the interlock specification:* Device panel section templates were generated interactively with edd and exported to .adl (ASCII) files. From the EPICS device RDB, ASCII meta-files were generated with the interlock information (see 2.3). A Perl script combined these meta-files and the section templates into complete device panel .adl files. The .adl files were converted with EPICS utilities into binary .dl files for the dm display manager. The meta-files could express only limited interlock complexity, but 97% of the ISAC vacuum device panels could be produced this way and have been in production use for the last 15 months. A few confusing situations arose when RDB specifications were out of synch with the ladder program and it was decided to move on to the next stage.

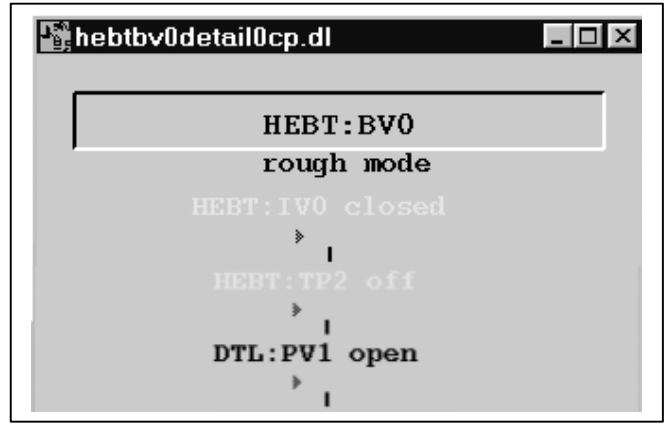

Figure 5: Automatically generated interlock detail panel

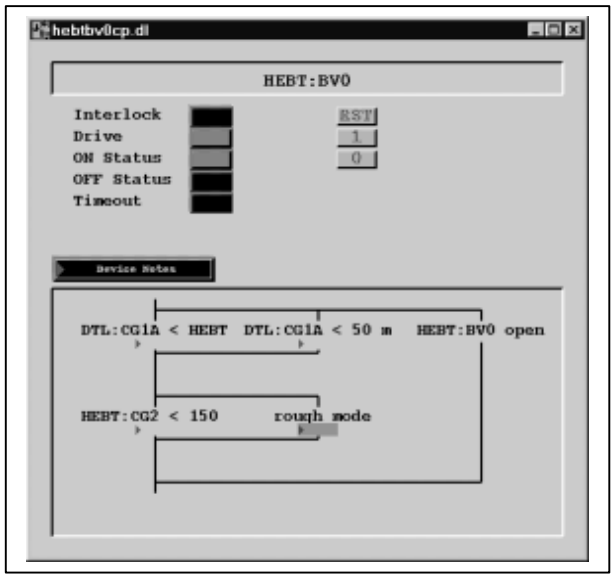

Figure 4: Automatically generated device control panels showing the interlocks of Fig. 1

c) *Automatic generation from the ladder program:* The interlock checking tool (see previous paragraph) had progressed at this stage to the point that it could easily be extended to generate meta-files that capture all the information in the interlock summary logic. Perl modules were written, which use these meta-files and generate device panel .adl files for all ISAC vacuum and ion source device classes. Also generated are detail panels for interlocks, which are only summarized on the device panel. This approach guarantees now absolute synchronization between the PLC ladder program and the EPICS device panels. "Hyperlink" buttons were added to each interlock condition, allowing quick system-wide call-up of device panels for offending devices. Figure 4 shows an automatically generated device panel displaying the interlocks shown in Figure 2, an automatically generated interlock detail panel, callable from the "rough mode" hyperlink in the device panel, is shown in Figure 5

## 3 SUMMARY

The automation of the interlock checking and interlock visualization improved the quality of the ISAC control system considerably. In addition it saves the controls group members from spending a lot of time on repetitive and error prone tasks. It remains to extend the automatic panel generation to the beam optics and beam diagnostics sub-systems. This should be less challenging as the interlocks are usually simple and all pertaining information is contained in the EPICS databases.

## ACKNOWLEDGEMENTS

The authors gratefully acknowledge the help of M. Leross who undertook the arduous task of checking the PLC interlocks, thereby finding the bugs in our utilities.

## REFERENCES

[1] R. Keitel et al., 'Control System Prototype for the ISAC Radioactive Beam Facility at TRIUMF', ICALEPCS97, Beijing, November 1997, p. 372

[2] R. Keitel, et al., 'Design and Commissioning of the ISAC Control System at TRIUMF', ICALEPCS99, Trieste, October 1999, p. 674